\documentstyle[12pt,aps]{revtex}
\hoffset=- 0.5 truecm
\def\R{ {\rm R \kern -.38cm I \kern .15cm}}
\def\C{ {\rm C \kern -.15cm \vrule width.5pt \kern .12cm}}
\def\Z{ {\rm Z \kern -.27cm \angle \kern .02cm}}
\def\N{ {\rm N \kern -.26cm \vrule width.4pt \kern .10cm}}
\def\1{{\rm 1\mskip-4.5mu l} }

\newcommand{\be}{\begin{equation}}
\newcommand{\ee}{\end{equation}}

\def\noi{\noindent}

\begin{document}
\baselineskip=20 pt

\begin{center}
{\Large \bf Pressure and Stress Tensor} \par \vskip 3 truemm
{\Large \bf in a Yukawa Fluid} \par \vskip 1 truecm
{\bf B. Jancovici}\footnote{Laboratoire de Physique Th\'eorique,
Universit\'e de Paris-Sud, B\^atiment 210, 91405 Orsay, France (Unit\'e
Mixte de Recherche n$^{\circ}$ 8627 - CNRS).\\ 
\noi e-mail: Bernard.Jancovici@th.u-psud.fr.}
\end{center}
\vskip 2 truecm
\begin{abstract}
Systems of particles interacting through a screened Coulomb potential of
the Debye-Yukawa form are considered. The pressure is obtained from the
stress tensor of the field corresponding to the Yukawa interaction, by a
suitable statistical average. This approach is especially appropriate
for systems living in a curved space. In a curved space, a self
contribution to the pressure appears, and it is essential to take it
into account for retrieving a correct pressure when the Yukawa
interaction tends to the Coulomb interaction.
\end{abstract}

\vspace{1 truecm}

 
\vskip 2 truecm

\noindent LPT Orsay 00-53 \par
\noindent June 2000 \par

\newpage

\noi {\bf 1. INTRODUCTION} 

\noi In a recent paper \cite{J} it has been discussed how the pressure
in a Coulomb fluid can be obtained by a suitable statistical average of
the Maxwell stress tensor. The present paper is an extension of these
calculations to the case of a fluid made of particles with Yukawa
interactions. These Yukawa fluids have been recently reviewed
\cite{SM,CG}. They are of interest because they are often used as
simplified models for some complex Coulomb fluids (for instance dusty
plasmas): one species $S$ of charged particles is given a microscopic
description, while the other charged particles are taken into account
only through the screening that they cause to the interaction between
the $S$-particles. One is left with one species interacting through a
screened Coulomb potential which is assumed to be of the Debye-Yukawa
type. In the following, this model will be considered for its own sake.  
In particular, the case when the fluid is in a curved space (of interest
both for numerical simulations and for looking at the curvature
effects) will be investigated.

To start with, the system under consideration is, in $\R^3$, a
gas of point-particles of number-density $n$, each of charge $q$, with
a pair interaction $v(r)=q^2 G(r)$ depending on the distance $r$, where
$G(r)$ has the Yukawa form
$$G(r)=\frac{\exp(-\alpha r)}{r} \eqno(1.1)$$

\noi which is the Green function of the Helmholz equation.From the
thermodynamic expression for the pressure 
$$P=-\partial F/\partial V \eqno(1.2)$$

\noi where $F$ is the free energy and $V$ the volume, one can derive
\cite{Han} the standard virial expression for the excess (over ideal)
pressure:  
$$P_{\mathrm{ex}}=-\frac{1}{6}n^2 q^2\int r\frac{dG}{dr}g(r)d{\bf r}
\eqno(1.3)$$ 

\noi where $g(r)$ is the pair distribution function (in the present
paper, integrals without domain specification are meant to be extended
to the whole manifold). In Section 2, it
will be shown how (1.3) can be alternatively 
derived from the stress tensor of the field which carries
the Yukawa interaction. In Section 3, the above considerations will be
extended to the case of a system confined on the surface $S^3$ of an
hypersphere. Section 4 deals with the two-dimensional analogs of these
systems. The limiting case of Coulomb systems is considered in Section
5; the case of a pseudosphere (a surface of constant negative curvature)
is revisited in Section 6. The results are discussed in Section 7. 
 
\noi {\bf 2. YUKAWA GAS IN} $\mathbf{\R^3}$

In terms of the Yukawa field $\phi ({\bf r})$ created by some charge
distribution, the energy density is 
$(1/8\pi)[(\bbox{\nabla}\phi)^2 +\alpha^2 \phi^2]$ and the
corresponding stress tensor is \cite{IZ}
$$T_{\mu\nu}=\frac{1}{4\pi}[\partial_{\mu}\phi \partial_{\nu}
\phi -\frac{1}{2} \delta_{\mu\nu}((\bbox{\nabla}\phi)^2
+\alpha^2\phi^2)] \eqno(2.1)$$

\noi where the Greek indices label the three Cartesian axes $x,y,z$. In
terms of the particle positions ${\bf r}_i$,
$$\phi({\bf r})=q\sum_i G(|{\bf r}-{\bf r}_i|) \eqno(2.2)$$   

The excess pressure is the negative of the statistical average of any 
diagonal element, say $-<T_{xx}>$, at some point which can be
chosen as the origin. Following the same steps as in ref.\cite{J},
one can decompose the excess pressure into nonself and self
parts. With the rotational symmetry taken into account, the nonself
part can be written as
$$P_{\mathrm{nonself}}=\frac{n^2 q^2}{24\pi}\int d{\bf r}_1 d{\bf r}_2      
[\bbox{\nabla}G(r_1)\cdot \bbox{\nabla}G(r_2) 
\: +3\alpha^2G(r_1)G(r_2)]g(|{\bf r}_2-{\bf r}_1|) \eqno(2.3)$$

\noi Writing each $G$ as the Fourier transform of $4\pi/(k^2+\alpha^2)$,
using as integration variables ${\bf r}_1$ and 
${\bf r}={\bf r}_2-{\bf r}_1$,
and performing first the integration on ${\bf r}_1$ , one recovers (1.3).
As to the self part of $-<T_{xx}>$,    
$$P_{\mathrm{self}}=-\frac{nq^2}{8\pi}\int d{\bf r}
[\left(\partial_xG(r)\right)^2
-\left(\partial_yG(r)\right)^2-\left(\partial_zG(r)\right)^2-
\alpha^2\left((G(r)\right)^2] \eqno(2.4)$$

\noi it is a divergent integral (at small $r$), which can however be
regularized through the use of the same physical argument as in
ref.\cite {J}: for computing in an unambiguous way the force per unit
area across the $yOz$ plane, one must assume that no particle sits on
that plane. Thus, one removes from the integration domain in (2.4) a
thin slab $-\varepsilon <x<\varepsilon$ and takes the limit 
$\varepsilon \rightarrow 0$ at the end of the calculation. After some
algebra with Fourier transforms, this prescription gives
$P_{\mathrm{self}}=0$.

Thus , in $\R^3$,
$$P_{\mathrm{ex}}=P_{\mathrm{nonself}} \eqno (2.5)$$

\noi and the stress tensor approach simply reproduces the standard
virial formula (1.3).    

\noi {\bf 3. YUKAWA GAS IN} $\mathbf{S^3}$

We now consider a Yukawa fluid living on the three-dimensional
``surface'' $S^3$ of an hypersphere of radius $R$. The Yukawa potential
(1.1) now must be changed into the Green function of the Helmholtz
equation on $S^3$ which is \cite{CG}
$$G(\psi)=\frac{1}{R}\frac{\sinh\omega(\pi-\psi)}{\sin\psi\sinh\omega\pi}
\ \mbox{for}\ \alpha R>1 \eqno(3.1a)$$
$$G(\psi)=\frac{1}{R}\frac{\sin\omega(\pi-\psi)}{\sin\psi\sin\omega\pi}
\ \mbox{for}\ \alpha R<1 \eqno(3.1b)$$  

\noi where $\psi$ is the angular distance seen from the center of the
hypersphere ($R\psi$ is the geodesic distance) and
$\omega=|\alpha^2 R^2-1|^{1/2}$.

\noi {\bf 3.1.Pressure from the stress tensor}

The analog of (2.3) is 
$$P_{\mathrm{nonself}}=\frac{n^2 q^2}{24\pi }\int dV_1 dV_2      
[\bbox{\nabla}_0 G(\psi_{01})\cdot\bbox{\nabla}_0 G(\psi_{02}) 
\: +3\alpha^2 G(\psi_{01})G(\psi_{02})]g(\psi_{12}) \eqno(3.2)$$

\noi where 0 is the arbitrary point at which $P_{\mathrm{nonself}}$
is evaluated, $dV_i$ is a volume element around point $i$, and
$R\psi_{ij}$ is the geodesic distance between points $i$ and $j$. After
a calculation described in Appendix A, (3.2) becomes
$$P_{\mathrm{nonself}}=-\frac{1}{6}n^2 q^2\int R\frac{\partial G(\psi)}
{\partial R}g(\psi)dV \eqno(3.3)$$ 

\noi where $G(\psi)$ must actually be regarded as a function of two
independent variables: the angular distance $\psi$ and the hypersphere
radius $R$, as seen on (3.1). In (3.3), the volume element is 
$dV=4\pi R^3\sin^2 \psi d\psi$. This result (3.3) is the generalization
of (1.3) to a curved space (the hypersphere). It should be noted that
now the distance $r$ has been replaced by the radius of curvature
$R$. Only in the flat system limit does $G(\psi,R)$ become a function of
the sole variable $r=R\psi$ and $R\partial G/\partial R=rdG/dr$.

Furthermore, now, the stress tensor approach provides another
contribution to $P_{\mathrm{ex}}$: the  
properly regularized part $P_{\mathrm{self}}$ does not vanish in the
$S^3$ case. Indeed, the analog of the integral (2.4) can be split into
two pieces $P_0$ and $P_1$ corresponding to the contributions of
geodesic distances to the origin smaller and larger than $R\psi_0$,
respectively. Only $P_0$ is divergent (at the origin) and must be
regularized by the same prescription as in the $\R^3$ case. It is
convenient to choose an infinitely small $\psi_0$. Then, the regularized
$P_0$ can be computed by using the small-$\psi$ form of (3.1), which is
just the Coulomb potential in $\R^3$. This calculation has been done in
ref.\cite{J} with the result 
$$P_0=-\frac{nq^2}{6R\psi_0}+O(\psi_0) \eqno(3.4)$$ 

\noi As to $P_1$, with the rotational symmetry taken into account, it
can be written as
$$P_1=\frac{nq^2}{24\pi}\int_{\psi>\psi_0}\left[\left(\frac{1}{R}
\frac{dG}{d\psi}\right)^2 +3\alpha^2 G^2\right]dV \eqno(3.5)$$ 

\noi This integral (3.5) is computed in Appendix A, and the final
result is 
$$P_{\mathrm{self}}=P_0+P_1=\frac{1}{6}nq^2
\left[\frac{\pi\alpha^2R}{\sin^2\pi\omega}
-\frac{\cot\pi\omega}{R\omega}\right]\ \mbox{for}\ \alpha R<1
\eqno (3.6)$$

\noi For $\alpha R>1$, $\omega$ must be replaced by $i\omega$ in (3.6).

Thus the stress tensor approach gives a total excess pressure
$P_{\mathrm{ex}}$ which is the sum of the nonself part (3.3) and the
self part (3.6). While the nonself part is the analog of the total
excess pressure (1.3) in $\R^3$, there is now in $S^3$ an additional
non-vanishing self term (3.6). 

\noi {\bf 3.2.Pressure from the free energy}   

Another way of defining and computing the pressure in the case of $S^3$
is to use the standard equation (1.2). If the total potential energy is 
$$W=q^2\sum_{i<j}G(\psi_{ij}) \eqno (3.7)$$
 
\noi the excess free energy is given by 
$$\beta F_{\mathrm{ex}}=-\ln\left\{\frac{1}{V^N}\int dV_1\ldots dV_N
\exp\left[-\beta \sum_{i<j}G(\psi_{ij})\right]\right\} \eqno(3.8)$$

\noi where $\beta$ is the inverse temperature and $N$ the number of
particles. From (1.2), where here $V=2\pi^2R^3$, one finds for the excess
pressure the value (3.3) which was the nonself part in the stress
tensor approach.

The self part (3.6) appears only if one includes in the total potential
energy the self-energies of the particles. Up to some (infinite)
volume-independent additive constant, the self-energy of a particle is
obtained in Appendix A as 
$$e_{\mathrm{self}}=-\frac{1}{2}q^2
\frac{\omega}{R}\cot\pi\omega \ \mbox{for}\ \alpha R<1 \eqno (3.9)$$

\noi If one adds to the free energy the self term
$Ne_{\mathrm{self}}$, one does obtain from (1.2) the additional self
contribution (3.6) to the pressure.  

These results will be discussed in Section 7.

\noi {\bf 4.TWO-DIMENSIONAL YUKAWA FLUIDS}

In this Section, the above considerations are adapted to the case of
two-dimensional systems. These two-dimensional systems are
toy models of theoretical interest. It should be
remembered that, in many experimental situations of charged particles
confined on a surface, the interaction nevertheless is the
three-dimensional Coulomb potential rather than the two-dimensional
interactions which are considered here.

\noi {\bf 4.1.Yukawa gas in} $\mathbf{\R^2}$

In $\R^2$, the Green function of the Helmholz equation is
$$G(r)=K_0(\alpha r) \eqno(4.1)$$

\noi where $K_0$ is a modified Bessel function and (1.3) is replaced by
$$P_{\mathrm{ex}}=-\frac{1}{4}n^2 q^2\int r\frac{dG}{dr}g(r)d{\bf r} 
\eqno(4.2)$$ 

In the field approach, the energy density is
$(1/4\pi)[(\bbox{\nabla}\phi)^2 +\alpha^2 \phi^2]$ and the corresponding
stress tensor is
$$T_{\mu\nu}=\frac{1}{2\pi}[\partial_{\mu}\phi \partial_{\nu}
\phi -\frac{1}{2} \delta_{\mu\nu}((\bbox{\nabla}\phi)^2
+\alpha^2\phi^2)] \eqno(4.3)$$

\noi where the Greek indices now label two Cartesian axes $x,y$.

When the rotational symmetry is taken into account, the analog of (2.3)
has the simpler form
$$P_{\mathrm{nonself}}=\frac{n^2 q^2}{4\pi}\alpha^2 \int d{\bf r}_1 
d{\bf r}_2 G(r_1)G(r_2)g(|{\bf r}_2-{\bf r}_1|) \eqno(4.4)$$

\noi (in two dimensions, the derivatives of $G$ cancel out). As in
the three-dimensional case, the expression (4.4) of 
$P_{\mathrm{nonself}}$ can be brought to the form (4.2) while the
properly regularized $P_{\mathrm{self}}$ vanishes. Thus, 
$P_{\mathrm{ex}}=P_{\mathrm{nonself}}$, and the stress
tensor approach reproduces (4.2). 

\noi {\bf 4.2.Yukawa gas in} $\mathbf{S^2}$

As shown in Appendix B, the Green function of the Helmholz equation on
a sphere $S^2$ of radius $R$ is
$$G(\theta)=-\frac{\pi}{2\sin\nu\pi}P_{\nu}(-\cos\theta) \eqno(4.5)$$

\noi where $\theta$ is the angular distance, $P_{\nu}$ is a Legendre
function, and 

\noi $\nu=(1/2)[-1+(1-4\alpha^2 R^2)^{1/2}]$ (if $2\alpha R>1$, $\nu$ is
complex).  

The analog of (4.4) is
$$P_{\mathrm{nonself}}=\frac{n^2 q^2}{4\pi}\alpha^2 \int dS_1
dS_2 G(\theta_{01})G(\theta_{02})g(\theta_{12}) \eqno(4.6)$$

\noi where $dS_i$ is an area element
around point $i$. After a calculation similar to the one for $S_3$ in 
Appendix A, (4.6) becomes the analog of (3.3)
$$P_{\mathrm{nonself}}=-\frac{1}{4}n^2 q^2\int R\frac{\partial G(\theta)}
{\partial R}g(\theta)dS \eqno(4.7)$$

\noi ($G(\theta)$ depends also on $R$ through $\nu$). When 
$P_{\mathrm{self}}$ is split into the contributions $P_0$ ($P_1$) of
geodesic distances smaller (larger) than $R\theta_0$, both parts remain
finite as $\theta_0\rightarrow 0$. The regularized $P_0$ is the same as
for a plane two-dimensional Coulomb system\cite{J}\ ($P_0=-n q^2 /4$)
and 
$$P_1=\frac{n q^2}{4\pi}\alpha^2 \int [G(\theta)]^2 dS \eqno (4.8)$$ 

\noi The integral in (4.8) involves $\int_{-1}^1 [P_{\nu}(x)]^2 dx$ 
which is tabulated in ref.\cite{GR}, and one obtains 
$$P_{\mathrm{self}}=P_0+P_1=\frac{n q^2}{4}\left\{-1+\frac{\alpha^2 R^2}
{2\nu +1} \left[\frac{\pi^2}{\sin^2 \pi\nu}-2\psi'(\nu+1)\right]\right\}
\eqno (4.9)$$
\noi where $\psi'$ is the derivative of the psi function (the psi
function being the logarithmic derivative of the gamma function).

The pressure can also be derived from the thermodynamic relation (1.2) 
(with the volume $V$ replaced by the sphere area $S$). Again, if only the
two-body interactions $q^2 G(\theta_{ij})$ are used for defining the free
energy $F$, only $P_{\mathrm{nonself}}$ is obtained, and, for 
$P_{\mathrm{self}}$ to appear, it is necessary to add to the free energy
the self-energies of the particles. Each particle is found (see Appendix
B) to have, up to some (infinite) volume-independent additive constant,
the self-energy 
$$e_{\mathrm{self}}=\frac{q^2}{2}\left[\ln R-\psi(\nu+1)-
\frac{\pi}{2}\cot\pi\nu\right] \eqno(4.10)$$

\noi This gives to the pressure the self contribution (4.9).

\noi {\bf 4.3.Yukawa gas in a pseudosphere}

The pseudosphere is a surface of constant negative curvature. Since it
is infinite, on it it is possible to study systems which are both
infinite and curved.

Let $a$ be the ``radius'' of the pseudosphere, such that the Gaussian
curvature is $-1/a^2$ (instead of $1/R^2$ on a sphere) and let $a\tau$
be the geodesic distance (instead of $R\theta$ on a sphere). As shown in
Appendix C, the Green function of the Helmholtz equation now is
$$G(\tau)=Q_{\nu}(\cosh \tau) \eqno(4.11)$$

\noi where $Q_{\nu}$ is a Legendre function of the second kind and

\noi $\nu =(1/2)[-1+(1+4\alpha^2 a^2)^{1/2}]$. 

The nonself pressure is given by (4.6) where $G$ now is $G(\tau)$ as
given in (4.11). After a calculation similar to the one in Appendix A,
one finds the analog of (4.7):
$$P_{\mathrm{nonself}}=-\frac{1}{4}n^2 q^2\int a\frac{\partial G(\tau)}
{\partial a}g(\tau)dS \eqno(4.12)$$

\noi As to the self pressure, its part $P_0$ is unchanged
($P_0=-nq^2/4$) while $P_1$ is given by (4.8) where $G$ is $G(\tau)$.
The integral involves $\int_1^{\infty}[Q_{\nu}(x)]^2 dx$ which is
tabulated in ref.\cite{GR}, and one obtains
$$P_{\mathrm{self}}=P_0+P_1=\frac{n q^2}{4}\left[-1+2\alpha^2 a^2
\frac{\psi '(\nu +1)}{2\nu +1}\right] \eqno (4.13)$$
 
\noi {\bf 5. COULOMB LIMIT} 

The results of ref.\cite{J} for a Coulomb fluid, the one-component
plasma, can be retrieved in the limit $\alpha\rightarrow 0$ of the
Yukawa fluid. However, before taking that limit, one must add a
neutralizing background to the Yukawa fluid. The presence of a
background is taken into account by changing $g$ into $h=g-1$ in
(1.3),(3.3),(4.2),(4.7).  

\noi {\bf 5.1.Coulomb limit in} $\mathbf{\R^3}$

As $\alpha\rightarrow 0$, $G\rightarrow 1/r$ and (1.3) becomes
$$P_{\mathrm{ex}}=P_{\mathrm{nonself}}=\frac{1}{6}n^2 q^2 \int
\frac{1}{r}h(r)d{\bf r} \eqno(5.1)$$  

\noi i.e.the excess pressure is one third of the potential energy density.

\noi {\bf 5.2.Coulomb limit in} $\mathbf{S^3}$

As $\alpha\rightarrow 0$,
$$R\frac{\partial G}{\partial R}=-\frac{6}{\pi\alpha^2 R^3}
-\frac{1}{\pi R}[(\pi-\psi)\cot\psi-\frac{1}{2}]+o(1) \eqno(5.2)$$

\noi Since the system is finite,
$$n\int h(\psi)dV=-1 \eqno (5.3)$$

\noi and (3.3) becomes  
$$P_{\mathrm{nonself}}=\frac{1}{6}
n^2 q^2\int\frac{1}{\pi R}[(\pi-\psi)\cot\psi
-\frac{1}{2}]h(\psi)dV-\frac{nq^2}{\pi\alpha^2 R^3}+o(1) \eqno(5.4)$$ 
 
\noi Furthermore, as $\alpha\rightarrow 0$, (3.6) becomes
$$P_{\mathrm{self}}=\frac{nq^2}{\pi\alpha^2 R^3}-\frac{nq^2}{4\pi R}
+o(1) \eqno(5.5)$$ 

\noi Thus, one retrieves the result of ref.\cite{J} for the total excess
pressure in the Coulomb case $\alpha=0$:  
$$P_{\mathrm{ex}}=P_{\mathrm{nonself}}+P_{\mathrm{self}}=\frac{1}{6}
n^2 q^2\int\frac{1}{\pi R}[(\pi-\psi)\cot\psi-\frac{1}{2}]h(\psi)dV
-\frac{nq^2}{4\pi R} \eqno(5.6)$$ 

It should be noted that, in the limit $\alpha\rightarrow 0$,
$P_{\mathrm{nonself}}$ and $P_{\mathrm{self}}$ have opposite divergent 
terms $O(\alpha^{-2})$ which cancel each other in their sum 
$P_{\mathrm{ex}}$. Thus, it is essential to keep the self term for
retrieving the finite result (5.6).

\noi {\bf 5.3.Coulomb limit in} $\mathbf{\R^2}$ 

As $\alpha \rightarrow 0$, $K_0(\alpha r)=-\ln (\alpha r/2)-\gamma
+o(1)$ (where $\gamma$ is Euler's constant)\cite{GR} and
$rdG/dr\rightarrow -1$. Since the Coulomb fluid exhibits perfect
internal screening, i.e. 
$$n\int h(r)d{\bf r}=-1 \eqno(5.7)$$ 

\noi (4.2) gives the known explicit exact result \cite{HH}
$$P_{\mathrm{ex}}=P_{\mathrm{nonself}}=-\frac{nq^2}{4} \eqno(5.8)$$

\noi{\bf 5.3.Coulomb limit in} $\mathbf{S^2}$
 
As $\alpha \rightarrow 0$, $\nu=-\alpha^2 R^2 +o(\alpha^2)$,
$P_{\nu}(-\cos \theta)=1+2\nu\ln\sin(\theta /2)+o(\alpha^2)$ \cite{GR}, and
$R\partial G/\partial R=-(1/\alpha^2 R^2)+o(1)$. Since the system is
finite, 
$$n\int h(\theta)dS=-1 \eqno(5.9)$$

\noi and (4.7) becomes
$$P_{\mathrm{nonself}}=-\frac{n q^2}{4\alpha^2 R^2}+o(1) \eqno(5.10)$$

\noi while (4.9) becomes 
$$P_{\mathrm{self}}=\frac{n q^2}{4}\left[-1+\frac{1}{\alpha^2 R^2}
+o(1)\right] \eqno(5.11)$$

\noi Thus, one retrieves the result of ref.\cite{J}:
$$P_{\mathrm{ex}}=P_{\mathrm{nonself}}+P_{\mathrm{self}}
=-\frac{n q^2}{4} \eqno(5.12)$$

Here too, $P_{\mathrm{nonself}}$ and $P_{\mathrm{self}}$ have opposite
divergent terms $O(\alpha^{-2})$ which cancel each other.

\noi {\bf 5.4.Coulomb limit in a pseudosphere}

As $\alpha \rightarrow 0$, $\nu \rightarrow 0$, $G(\tau) \rightarrow
Q_0(\cosh \tau)=-\ln \tanh(\tau /2)$, and $a\partial G/\partial a
\rightarrow 0$. Thus $P_{\mathrm{nonself}}=0$. On the other hand, (4.13) 
becomes $P_{\mathrm{self}}=-n q^2/4$. Thus,
$$P_{\mathrm{ex}}=P_{\mathrm{self}}=-\frac{nq^2}{4} \eqno(5.13)$$

Here, the excess pressure entirely comes from the self part.  
 
\noi {\bf 6. VIRIAL EXPANSION IN A PSEUDOSPHERE}

In a pseudosphere, Jancovici and T\'ellez \cite{JT} have defined an
excess pressure by the virial expansion in powers of the density $n$
$$\beta p_{\mathrm{ex}}=\sum_{k=2}^{\infty}B_k n^k \eqno(6.1)$$

\noi Obviously, (6.1) does not agree with (5.13). The present Section is
an erratum to ref.\cite{JT}: (6.1) is \emph{not} a good definition of
the pressure. The considerations about the virial expansion in
ref.\cite{JT} should be replaced by what follows.

Since on a pseudosphere, as the size of a domain is increased, its
perimeter grows as fast as its area $S$, there is no unique
thermodynamic limit for the free energy per unit area $F/S$. A
reasonable definition of a bulk quantity is provided
by the usual virial expansion of \emph{the free energy}, obtained by
manipulations of the partition function: the nonself
part of the excess free energy per particle can be defined by
$$\beta f_{\mathrm{nonself}}=\sum_{k=2}^{\infty}\frac{B_k}{k-1}n^{k-1}
\eqno(6.2)$$

\noi where each virial coefficient $B_k$ is to be computed in the
infinite system limit (for eliminating the boundary effects) before the
sum is performed. 

 From this free energy (6.2), one can derive the nonself part of the
excess pressure by the standard relation
$P_{\mathrm{nonself}}=n^2\partial f_{\mathrm{nonself}}/\partial n$
provided one takes into account an unusual feature (which has been
disregarded in ref.\cite{JT}): the interaction law $G$ and therefore 
the excess free energy per particle $f_{\mathrm{ex}}$ depends on the
radius of curvature $a$ (the virial coefficients $B_k$ do depend on
$a$): it is convenient to consider $f_{\mathrm{ex}}$ as a function of
the two independent variables $n$ and $na^2$. The variation of density
in the above definition of the excess pressure can be considered as
associated to a variation of $a$ while the average number of particles
in some domain remains constant as the area of this domain varies
proportional to $a^2$. In other words, for defining the excess pressure,
the partial derivative of the excess free energy with respect to $n$
must be taken at constant $na^2$, and one must write more precisely 
$$P_{\mathrm{nonself}}=n^2\left(\frac{\partial f_{\mathrm{nonself}}}
{\partial n}\right)_{na^2} \eqno(6.3)$$.

\noi In the case of a Coulomb system of point particles, for dimensional
reasons, $f_{\mathrm{nonself}}$ depends on $n$ and $a$ only through the
combination $na^2$ ($B_k$ is proportional to $a^{2(k-1)}$), and (6.3)
gives $P_{\mathrm{nonself}}=0$.

Furthermore, self effects must be taken into account. For a Coulomb
system, with an interaction $G=-\ln\tanh(\tau/2)$, the self-energy of a 
particle of small radius $r_0=a\tau_0$ 
$$e_{\mathrm{self}}=f_{\mathrm{self}}=-\frac{q^2}{2}\ln\frac{r_0}{2a}=
-\frac{q^2}{4}\ln\frac{r_0^2 n }{4na^2} \eqno(6.4)$$

\noi does give a contribution to the pressure
$$P_{\mathrm{self}}=n^2\left(\frac{\partial f_{\mathrm{self}}}
{\partial n}\right)_{na^2}=-\frac{nq^2}{4} \eqno(6.5)$$

\noi in agreement with (5.13).
   
\noi {\bf 7. DISCUSSION}

In a flat space $\R^3$ or $\R^2$, the stress tensor approach
simply reproduces the expected pressure. There is no self contribution. 

The situation is more involved in the finite curved spaces $S^3$ and
$S^2$, and in the infinite pseudosphere. The very definition of the
pressure is not obvious. An operational definition would be the force
per unit area exerted on a wall bounding the fluid. However, when the
system is a finite one, such a pressure would depend on the position and
shape of this wall, and also, for a fixed curvature, the mere presence of a
boundary would change the size of the system. In the case of a
pseudosphere, a system with a boundary makes difficulties because its
perimeter grows as fast as its area. The stress tensor approach, which  
has been used here, has the advantage of defining a bulk pressure which
does not refer to any wall. But it raises a question: should the self
part be included or not in the definition of the pressure? One might be
tempted to follow the usual procedure of discarding  self effects.
However, as seen in Section 5, it is necessary to keep this self term
for retrieving a finite pressure in the Coulomb case. Thus, the nonself
and self parts of the pressure are somewhat entangled with each
other. For a flat system in $\R^3$ or $\R^2$, the (properly regularized)
self pressure vanishes, for a Yukawa system and thus in the Coulomb
limit. However, the
same pressure for a flat Coulomb system can be obtained by starting with
a Yukawa system (plus background) in $S^3$, $S^2$, or the pseudosphere,
going to the Coulomb  limit $\alpha \rightarrow 0$, and finally going to
the limit $R\rightarrow \infty$ or $a\rightarrow \infty$ of a flat
system; if this route is followed, it is mandatory to keep the self
part of the pressure.  

The entanglement of the nonself and self parts of the pressure is
especially apparent in the two-dimensional case. The result for a
flat Coulomb system, $P_{\mathrm{ex}}=-nq^2/4$ can be obtained by
starting with a Yukawa system on a sphere $S^2$, and taking the limits 
$\alpha \rightarrow 0$ and $R \rightarrow \infty$. These limits can be
taken in one order or in the opposite one, or even for a fixed value of 
$\alpha R$, giving the same final result for $P_{\mathrm{ex}}$. 
However, the separate contributions $P_{\mathrm{nonself}}$ and
$P_{\mathrm{self}}$ do depend on the way the limits are taken.
 
In a curved space, the pressure from the stress tensor approach
can also be retrieved from the usual thermodynamic definition (1.2)  
(the negative of the derivative of the free energy with respect to the
volume, or its two-dimensional analogs). If one deals from the start
with a Coulomb fluid in $S^3$ or $S^2$, there is some arbitrariness in
the definition of the self-energy
of a particle \cite{CL}, and only a ``reasonable'' choice allows to
retrieve \cite{J} the stress tensor pressure. In the case of a Yukawa
fluid, the self-energy (3.9) or (4.10) is sufficiently well-defined for  
the calculation of the pressure. Starting with a Yukawa fluid and going
afterwards to the Coulomb limit avoids the above mentioned
arbitrariness.

It should be noted that, in curved spaces, the stress tensor approach
defines the pressure as the response to a change of the radius
$R$ or $a$; the thermodynamic approach does the same.
However, as $R$ or $a$ changes, the interaction 
potential changes, and therefore the derivative $-\partial F/\partial V$
is \emph{not} taken at constant interaction potential. This is a bit
unusual! It has to be taken into account when using the virial expansion
which seems to exist in the pseudosphere.

The conclusion is that there is some arbitrariness in the definition of
the pressure in a curved space. However, the stress tensor approach, in
which there is some reason for including the self contribution, gives a
pressure which seems acceptable.
   
\noi {\bf APPENDIX A. FORMULAS IN} $\mathbf{S_3}$

For deriving (3.3) from (3.2), one first note that, since the integrand
in (3.2) depends only on the shape of the geodesic triangle formed by
points (0,1,2), the integration can be performed on another pair of
positions (0,2), rather than (1,2). By an integration by parts on 0, the
term   
$\bbox{\nabla}_0 G(\psi_{01})\cdot\bbox{\nabla}_0 G(\psi_{02})$
can be replaced by $-G(\psi_{01})\bigtriangleup_0 G(\psi_{02})$. Using
the Helmholtz equation with a point source
$$\left[-\bigtriangleup_0 +\alpha^2\right] G(\psi_{02})=
-4\pi\delta^{(3)}(\psi_{02}) \eqno(A.1)$$

\noi gives
$$P_{\mathrm{nonself}}=\frac{n^2 q^2}{24\pi}\left[4\pi\int dV_2
G(\psi_{12})g(\psi_{12})\ +2\alpha^2 \int dV_0 dV_2 G(\psi_{01})
G(\psi_{02})g(\psi_{12})\right] \eqno(A.2)$$

\noi Using the Dirac notation 
$$G(\psi_{ij})=<i|\frac{4\pi}{-\bigtriangleup +\alpha^2}|j> \eqno(A.3)$$

\noi where $i,j$ are positions, gives for the integral on 0 in (A.2)
$$\int dV_0 G(\psi_{10})G(\psi_{02})=
\int dV_0 <1|\frac{4\pi}{-\bigtriangleup +\alpha^2}|0>
<0|\frac{4\pi}{-\bigtriangleup +\alpha^2}|2>$$ 
$$= <1|\frac{(4\pi)^2}{(-\bigtriangleup +\alpha^2)^2}|2>
=-4\pi\frac{\partial G(\psi_{12})}{\partial (\alpha^2)} \eqno(A.4)$$

\noi Thus, with simpler notations $V$ and $\psi$ instead of $V_2$ and
$\psi_{12}$, (A.2) becomes
$$P_{\mathrm{nonself}}=\frac{n^2 q^2}{6}\int dV\left[
G(\psi)-\alpha\frac{\partial G(\psi)}{\partial\alpha}\right]g(\psi) 
\eqno(A.5)$$.

\noi Finally, since $RG(\psi)$ as defined in (3.1) depends on $\alpha$
only through $\alpha R$, $\alpha(\partial G/\partial\alpha)= 
\partial (RG)/\partial R$ and (A.5) gives (3.3).

For deriving (3.6) from (3.4) and (3.5), first one performs an
integration by parts which transforms (3.5) (where $dV=4\pi R^3 \sin^2
\psi d\psi$) into
$$P_1=\frac{nq^2}{24\pi}\left\{\int_{\psi>\psi_0}\left[-G(\psi)
\bigtriangleup G(\psi)+3\alpha^2 \left[G(\psi)\right]^2\right]dV
-\left[G(\psi)\frac{dG}{d\psi}4\pi R \sin^2 \psi\right]_{\psi=\psi_0}
\right\} \eqno(A.6)$$

\noi Since $G$ obeys the Helmholz equation and $\psi\neq 0$, in (A.6) 
the Laplacian $\bigtriangleup$ can be replaced by
$\alpha^2$. Furthermore, the integral remains convergent as $\psi_0
\rightarrow 0$ and can be extended to the whole hypersphere. With $G$
given by (3.1), this is an elementary integral on trigonometric or
hyperbolic functions. Adding the small-$\psi_0$ form of the last term of
(A.6), one finds
$$P_1=\frac{1}{6}nq^2\left[\frac{\pi\alpha^2 R}{\sin^2 \pi\omega}-
\frac{\cot \pi\omega}{R\omega}+\frac{1}{R\psi_0}+O(\psi_0)\right] 
\ \mbox{for}\ \alpha R<1 \eqno(A.7)$$

\noi Adding (3.4) and (A.7) gives (3.6). A similar calculation can be
done when $\alpha R>1$.

For computing the self-energy (3.9) of a particle pictured as a
spherical surfacic charge of radius $r_0=R\sin\psi_0$, in the
small-$\psi_0$ limit one can use the point-particle potential (3.1).
As $\psi_0 \rightarrow 0$
$$e_{\mathrm{self}}=\frac{q^2}{2}G(\psi_0)=\frac{q^2}{2}\left[
\frac{1}{r_0}-\frac{\omega\cot\pi\omega}{R}+O(\psi_0)\right]
\ \mbox{for}\ \alpha R<1 \eqno(A.8)$$

\noi Up to the volume independent term $q^2/2r_0$ (which becomes
infinite in the $\psi_0 \rightarrow 0$ limit), (A.8) does give
(3.9). For $\alpha R>1$, $\omega$ must be replaced by $i\omega$. 

\noi {\bf APPENDIX B. FORMULAS IN} $\mathbf{S_2}$

For finding the Green function (4.5), one notes that the Helmholtz
equation in $S_2$ (similar to (A.1) with $2\pi\delta^{(2)}$ instead of
$4\pi\delta^{(3)}$) reduces to the Legendre equation\cite{GR}, with the
variable $\cos\theta$ and the parameters
$\nu =(1/2)[-1+(1-4\alpha^2 R^2)^{1/2}],\ \mu=0$.The solution singular
at $\theta=0$ and regular at $\theta=\pi$ is the Legendre function 
$$P_{\nu}(-\cos\theta)=F(-\nu,\nu+1;1;\frac{1+\cos\theta}{2}) \eqno(B.1)$$  

\noi Indeed, at $\theta=\pi$ the hypergeometric function $F$ is regular,
and as $\theta \rightarrow 0$ its behavior\cite{E}
$$P_{\nu}(-\cos\theta)=\frac{2\sin\nu\pi}{\pi}\left[\ln\sin\frac{\theta}{2}
+\gamma +\psi(1+\nu)+\frac{\pi}{2}\cot\nu\pi \right]+o(1) \eqno(B.2)$$

\noi is\footnote[2]{There is a minor misprint in ref.\cite{E}. In
eq.(15) p.164 $\gamma$ must be replaced by $2\gamma$ for this
equation to agree with eq.(12) p.110.} such that (4.5) does behave like
$-\ln\theta$. 

For deriving the self-energy (4.10), one looks at the behaviour of 
$(q^2 /2)G(\theta_0)$, considering $r_0=R\theta_0$ as a small fixed 
particle radius. As $\theta_0 \rightarrow 0$, using (B.2) in (B.1)
gives (4.10).

\noi {\bf APPENDIX C. GREEN FUNCTION IN A PSEUDOSPHERE}

The Helmholtz equation in a pseudosphere of ``radius'' $a$ reduces to
the Legendre equation\cite{GR} with the variable $\cosh\tau$ and the
parameters $\nu =(1/2)[-1+(1+4\alpha^2 a^2)^{1/2}],\ \mu=0$. The
solution singular at $\tau=0$ and vanishing at infinity is
$Q_{\nu}(\cosh\tau)$. For small $\tau$, it does behave\cite{E} like 
$-\ln\tau$.   

\noi {\bf ACKNOWLEDGEMENT}

I am indebted to J.M.Caillol for stimulating discussions and a critical
reading of the manuscript.

\newpage

\noi 

\end{document}